\documentclass[
aps,
prl,
twocolumn,
superscriptaddress,
floatfix
longbibliography
]{revtex4-1}
\usepackage[dvipdfmx]{graphicx}
\usepackage{verbatim}
\usepackage{epsf}
\usepackage{natbib}
\usepackage{dcolumn}
\usepackage{bm}
\usepackage{color} 
\usepackage{ulem}
\usepackage{amsmath}
\usepackage{amssymb}
\usepackage{txfonts}
\usepackage{bigints}
\usepackage[
colorlinks=true,
citecolor=blue,
urlcolor=blue,
setpagesize=false]{hyperref}

\begin{document}

\title{
 Solution of the ``sign problem'' for the half filled  Hubbard-Holstein model
}
\author{Seher~Karakuzu}
\affiliation{International School for Advanced Studies (SISSA), Via Bonomea 265, 34136, Trieste, Italy} 

\author{Kazuhiro~Seki}
\affiliation{International School for Advanced Studies (SISSA), Via Bonomea 265, 34136, Trieste, Italy}
\affiliation{Computational Materials Science Research Team, RIKEN Center for Computational Science (R-CCS),  Hyogo 650-0047,  Japan}

\author{Sandro~Sorella}
\affiliation{International School for Advanced Studies (SISSA), Via Bonomea 265, 34136, Trieste, Italy} 
\affiliation{Computational Materials Science Research Team, RIKEN Center for Computational Science (R-CCS),  Hyogo 650-0047,  Japan}

\begin{abstract}
We show that, by an appropriate choice of auxiliary fields and exact integration of the phonon degrees of freedom, it is possible to define a  ''sign-free'' path integral 
for the so called Hubbard-Holstein model at half-filling.
 We use 
 a statistical 
 method, based on an accelerated and efficient Langevin dynamics,
for  evaluating all relevant correlation functions of the model. 
 Preliminary calculations at $U/t=4$ and $U/t=1$, for $\omega_0/t=1$, indicate 
 a quite extended region around $U \simeq {g^2 \over \omega_0}$ without either antiferromagnetic or charge-density-wave orders, separating two quantum critical points at zero temperature.
The elimination of the sign problem in  a model without explicit particle-hole symmetry may open new perspectives for strongly correlated models, even away from the purely attractive or particle-hole symmetric cases.
\end{abstract}

\date{\today}

\maketitle

\textit{Introduction: }
One of the most successful methods to obtain exact properties 
of strongly correlated models on a lattice 
is certainly the statistical  (Monte Carlo) method based on the 
evaluation of a corresponding path integral defined in imaginary time. 
In particular most successful applications are based on  
auxiliary fields $\sigma_{l}^j$ introduced for each site $j$ of the model 
and imaginary time slice $l$ of the path integral by means of  the so called ``Hubbard-Stratonovich'' transformation (HST)~\cite{sugar,Hirsch1983,me,Assaad2005}. Since the Hirsch seminal work in '85~\cite{Hirsch1985}, 
several models have been studied, and their phase diagrams have been solved numerically for large enough 
number $N$ of sites, in very particular cases when the so called sign problem does not affect the simulation of the corresponding partition function  $Z=\int \left[ d\sigma_{l}^j \right] W(\left\{\sigma_{l}^j\right\})$, that is evaluated by standard statistical methods, as long as $ W(\left\{\sigma_{l}^j\right\})\ge 0$. The first example was the Hubbard model in the square lattice, displaying a trivial phase diagram for the insulating antiferromagnetic phase, 
that turned out to be stable as soon as $U>0$. More recently the method was extended to the honeycomb lattice displaying a less trivial transition at a critical value $U_c$ between a semimetallic and an antiferromagnetic insulating or superconducting phase~\cite{Sorella1992,Sorella2012,Otsuka2016,Otsuka2018}. Other models are worth to be mentioned such as, 
the negative-$U$ model~\cite{Scalettar1989,Karakuzu2018}, the spinless fermions with 
repulsive nearest-neighbors interaction at half filling~\cite{Fulton2014,spinless_sign}, the 
Anderson impurity model at half filling~\cite{Hirsch1986,Feldbacher2004}, the Kondo-lattice model 
at half filling~\cite{Assaad1999,Capponi2001}, the Holstein-model~\cite{Hirsch1983_2} 
and several others.

Most of  the models so far solved without sign problem  are characterized by i) an explicit spin-independent attractive interaction and/or ii)  a particular particle-hole symmetry of the electronic degrees of freedom, implying that the corresponding weight $W(\{\sigma_{l}^j\})$ 
in the path-integral 
formulation can be written as the square of a quantity, and therefore positive. All these models  have been recently classified in Ref.~\cite{Wang2015,sign_class}.
For instance in the Hubbard model with $U>0$, the particle-hole transformation  \begin{equation}  \label{phole}
c^\dag_{i\downarrow} \to c_{i\downarrow}
\end{equation}
(  $c^\dag_{i\downarrow} \to -c_{i\downarrow}$) for sites $i$ in the A (B) sublattice of a bipartite lattice, 
maps the positive-$U$ model to the 
negative-$U$ one with equal number of spin-up and spin-down electrons, where $c^\dag_{j,\sigma}(c_{j,\sigma})$ creates (destroys) a fermion with spin $\sigma = \uparrow,\downarrow$ at a given site $j$.
In such case the weight factorizes into two independent and identical contributions for different spins, thus $W(\{ \sigma_l^j \} )>0$.

The Hubbard-Holstein Hamiltonian is one of the simplest model describing the competition between an attractive interaction 
mediated by an optical phonon and the strong electron repulsion, defined by the Hubbard  $U$, acting when two electrons of opposite spins occupy the same site.
The Hubbard-Holstein model represents the key model to understand how the retarded interaction mediated by phonons can circumvent the strong 
electron-electron repulsion and give raise to superconductivity. 
It may be relevant not only to understand standard electron-phonon superconductivity, but also 
the high-temperature one, because the isotope effect has been clearly detected~\cite{Hofer2000} in cuprates, and the 
so called kinks observed in photoemission experiments~\cite{Lanzara2001} clearly indicate the role of phonons, even in these strongly 
correlated materials.

The phase diagram of the model has been studied using several techniques such as, Gutzwiller approximation \cite{fabrizio},  variational Monte Carlo (VMC)~\cite{Ohgoe2017,Karakuzu2017}, dynamical mean-field theory (DMFT) \cite{Millis,capone,Bauer2010}, finite-temperature determinant quantum Monte Carlo (DQMC) \cite{Nowadnick2012,Johnston2013,Weber}, also in 1D~\cite{Assaad_1dHM}, but no unbiased zero temperature calculation is known in 2D.

In the present work we are able to establish ground-state benchmark results in the thermodynamic limit for this model, and some aspects of its zero-temperature phase diagram, by using 
a determinantal method, which, as we are going to show, is not vexed by 
the so called ``sign problem''.

\textit{Model and Method: }
The Hubbard-Holstein model is defined by the following Hamiltonian:
\begin{eqnarray}
{\cal H} &=&{\cal H}_{\cal K} + {\cal H}_{\cal V},\nonumber \\
  {\cal H}_{\cal K} &=&  K +\frac{\omega_0}{2} \sum_{j} \hat{P}_{j}^2, \nonumber \\
  {\cal H}_{\cal V} &=&  \frac{U}{2} \sum_j (n_j-1)^2  + g \sum_j \hat{X}_{j}  (n_j-1)+  \frac{\omega_0}{2} \sum_{j}  \hat{X}_{j}^{2} \nonumber \\
 &=&  \frac{U}{2} \sum_j (n_j-1 + { g \over U} \hat{X}_{j})^2   +  \frac{\omega_0-g^2/U}{2} \sum_{j}  \hat{X}_{j}^{2},
\end{eqnarray}
where $K =  -t \sum_{\langle i,j \rangle , \sigma} 
c^\dag_{i,\sigma} c_{j,\sigma} +{\rm H. c.}$, $n_{j \sigma} = c^\dag_{j,\sigma}c_{j,\sigma}$ indicates the electron number with spin $\sigma$ at the  site $j$,
and $n_j=\sum_{\sigma} n_{j \sigma}$. $t$ is the hopping integral, $U$ is the repulsive electron-electron interaction, whereas, $\omega_0$ is the phonon frequency, $g$ is the electron-phonon coupling term and, $\hat{X}_j$ and $\hat{P}_j$ are phonon position and momentum degrees of freedom, respectively. 

At finite inverse temperature $\beta$
the partition function of an electronic  system described by the Hamiltonian $\mathcal{H}$ is given by:
\begin{equation}\label{eq:pf}
  \mathcal{Z} = {\rm Tr} [e^{-\beta \cal H}]
  = {\rm Tr} [(e^{-\Delta \tau \cal H})^{T}],
\end{equation}
where  $\Delta \tau = \beta /T$ and the symbol $ {\rm Tr}[ O]$ associates a number to any operator $O$ and is  defined by: 
\begin{eqnarray}
{\rm Tr}[ O] =\left\{ 
\begin{array}{cc}
 {\rm Trace} [ O] & {\rm for ~the~ standard~case} \\
 \langle \Psi| O |\Psi \rangle  & \text{  for ~the~ projection~case}
\end{array} \right.
\end{eqnarray}
where $\Psi$ is a chosen trial function, that may be conveniently introduced for evaluating the trace 
(up to an irrelevant constant) in the zero-temperature limit, as long as $\Psi$ has a non zero overlap with the ground state of ${\cal H}$.
The latter  case is  known as zero-temperature projection, that is adopted in all the forthcoming calculations. 
However, for the sake of generality, we derive the method in the general case, as it is defined also 
within the more conventional finite-temperature scheme.

The imaginary-time propagator $e^{-\beta {\cal H}}$ can be written after Trotter decomposition as:
\begin{equation}\label{eq:trott}
  e^{-\Delta \tau \cal H} = 
  e^{-\Delta \tau {\cal H}_{\cal V} } e^{-\Delta \tau {\cal H}_{\cal K} }  + O(\Delta \tau^{2}), 
\end{equation}
In order to derive the path integral, the phonon degrees of freedom, introduced for each site $j$ and time slice $l$,  are dealt as in a conventional Feynmann path integral where the phonon positions 
$X^j_l$ are changed at different time slices, just due to the phonon kinetic 
energy $ { \omega_0 \over 2} \sum\limits_j \hat{P}_j^2$, with associated matrix elements:
\begin{equation}
 \langle X^j_{l+1} | \exp{\left(-{\Delta \tau \omega_0 \over 2}  \hat{P}_j^2 \right)} | X^j_l \rangle \propto \exp{\left[ - { 1 \over 2 \omega_0 \Delta \tau} ( X^j_{l+1}-X^j_l)^2 \right]}
\end{equation}
After that the operators $\hat X_j$ turn onto classical real variables $X_j^l$ 
to be integrated from $-\infty$ to $\infty$ 
in the corresponding path integral. For the remaining interaction term in ${\cal H}_{\cal V}$ we can use a properly chosen 
HST coupled to the operator $n_j-1 + {g \over U} X_l^j$, namely:
\begin{equation}\label{eq:HFcharge}
e^{-\frac{U \Delta\tau}{2} (n_{j} -1 + {g \over U} X_l^j)^2} = \int_{-\infty}^{\infty} \frac{d\sigma_l^j}{\sqrt{2\pi}} e^{-\frac{1}{2}(\sigma_l^j)^2 +i \sqrt{U \Delta \tau} \sigma_l^j (n_{j} -1 + {g \over U} X^j_l)} 
\end{equation}
where $i$ is the imaginary constant,  $\sigma_j^l$ are indicating the auxiliary fields in the $l^{\rm th}$ time slice.
We thus get that the partition
function 
can be expressed as a $2N\times T$ dimensional
integral over the classical real variables $\sigma_{l}^j$ and $X_{l}^j$:

\begin{eqnarray} \label{path}
\mathcal{Z} & =& \bigintss 
\left[ dX \right] \bigintss \left[ d\sigma \right]   \exp \left[  -{1 \over 2} \sum \limits_{l,m,j} \left( A_{l,m} X^j_l X^j_m +\delta_{l,m} (\sigma^j_l)^2 \right) \right]   \nonumber \\ 
& \times & {\rm Tr} \prod\limits_{l=1}^T \left\{ \exp\left[ \sum\limits_j \left( i \sqrt{U \Delta\tau} \sigma^j_l \right) (n_j-1 +{g \over U} X^j_l ) \right] \exp( - \Delta\tau K) \right\}  \nonumber \\
\end{eqnarray}
where the product of non-commuting operators is meant from left to right with increasing $l$, $\left[ dX \right] =\prod\limits_{j,l} dX_l^j$, 
$ \left[ d\sigma \right] = \prod\limits_{j,l} d\sigma_l^j$, and:
\begin{equation}
A_{l,m} = { 1 \over \omega_0 \Delta\tau } \left[ 2 \delta_{l,m} -\delta_{l,m+1} -\delta_{m,l+1} \right] +\Delta\tau (\omega_0- g^2/U) \delta_{l,m}.
\end{equation}
Here the boundary conditions for the phonon fields are $X^j_{l+T}=X^j_l$ 
for the standard finite temperature case (periodic in imaginary time) 
and $X^j_0=X^j_{T+1}=0$ for the projection case (open boundaries in imaginary time, within appropriate trial function $\Psi$ \cite{note_matA}). 
In the path integral the dependence of the action on the fields $X_l^j$ is just quadratic and 
determined by the matrix $A$.
After simple inspection,
the eigenvalues  of $A$ are given by 
\begin{equation} \label{diagA}
E_n= {1 \over \omega_0 \Delta\tau} [ 2 -2 \cos (\omega_n\Delta\tau)]  + \Delta\tau (\omega_0 - g^2/U) 
\end{equation}
where $\omega_n \Delta\tau = {2 \pi n \over T} $ 
($\omega_n \Delta \tau={n \pi \over T+1}$) for the finite temperature (projection) case and $n=1,\cdots T$.
Therefore  this matrix $A$  is positive definite 
($E_n>0 \forall n$) 
if $U> g^2/\omega_0$ and all the integrals in $ \{ X_l^j\}$ can be 
carried out before the $\{\sigma_l^j \}$ ones, as they are certainly 
converging:
\begin{eqnarray} 
\mathcal{Z} &\propto& \bigintss \left[ d \sigma \right]  \exp \left[  -{1 \over 2} \sum \limits_{l,m,j} \left( P_{l,m} 
 \sigma^j_l  \sigma^j_m \right) \right] \nonumber \\
&\times& {\rm Tr} \prod\limits_l \left\{ \exp\left[i \sqrt{U \Delta \tau}  \sum\limits_j   \sigma^j_l (n_j-1) \right] \exp( - \Delta \tau K) \right\} \nonumber \\  \label{wght}
\end{eqnarray}
where: 
\begin{equation}
P_{l,m} = \delta_{l,m} +  {\Delta\tau  g^2 \over U}  
\left[ A^{-1}\right]_{l,m}.
\end{equation}
is real and providing a positive weight in Eq.~(\ref{wght}).
On the other hand the remaining part contributing to the path integral weight $W(\{ \sigma_l^j \})$, and resulting from the $ {\rm Tr}$ operation is certainly positive, because 
the spin-up and spin-down contributions factorize and, 
after the particle-hole transformation in Eq.~(\ref{phole}),
turn out to be  complex-conjugate factors [with appropriate $\Psi$ in the projection case]. Thus we have finally determined a path 
integral for the Hubbard-Holstein model with a positive  
real weight $W(\{ \sigma_l^j \} ) >0$ for $U> g^2/\omega_0$.

\begin{figure}
  \begin{center}
    \includegraphics[width=.9\columnwidth]{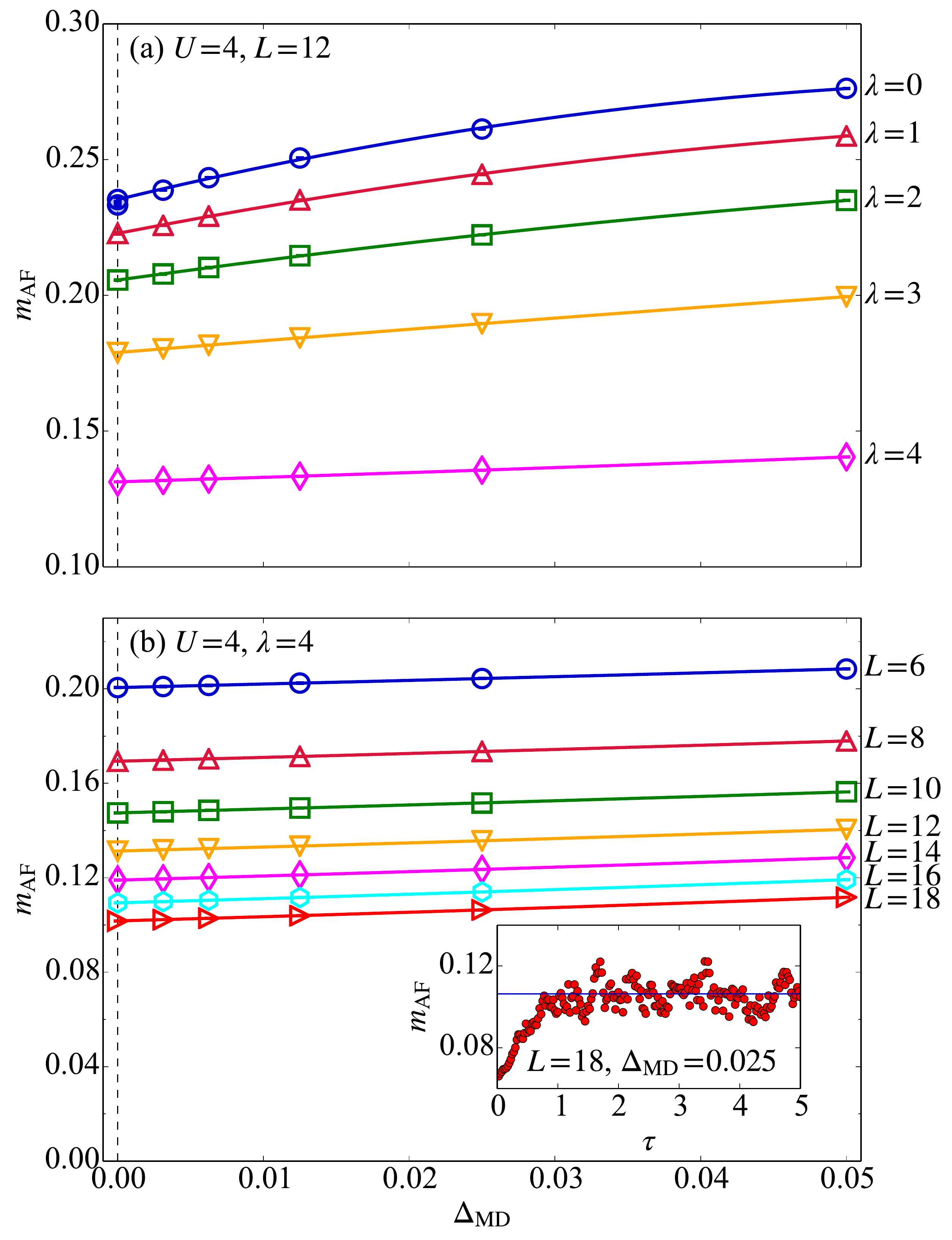}
    \caption{
      \label{fig1}
      Antiferromagnetic order parameters $m_{\rm AF}$ at $U/t=4$ 
      as a function of the 
      MD time step $\Delta_{\rm MD}$ 
      (a) on a 12 $\times$ 12 cluster at $\beta t=12$ and different values of the electron-phonon coupling strength $\lambda$. 
      The $\lambda=\Delta_{\rm MD}=0$ result (filled circle) in (a)
      is obtained with the standard Monte Carlo algorithm for the Hubbard model, that is clearly consistent with the MD data, extrapolated to $\Delta_{\rm MD} \to 0$.
      (b) same as (a) for $\lambda=4$ with various system sizes 
      $N=L\times L$, with $\beta t=L$.       
      The inset  shows the equilibration of $m_{\rm AF}$ to its average value (blue line)  for  the largest cluster as a function of the MD time $\tau$. 
    }
  \end{center}
\end{figure}

A standard approach to evaluate correlation functions defined by the 
partition function $\mathcal{Z}$ is  the  Monte Carlo (MC) method.
Unfortunately, the standard technique with local updates is very inefficient in this case, due to the difficulty to sample the stiff harmonic part. A better method was recently  introduced~\cite{Johnston2013}, including  global updates of the phonon fields. 
Global updates are numerically very demanding as  they require the computation of several determinants from scratch. Here, in order to define an efficient sampling,  we will use the first-order accelerated Langevin dynamics~\cite{Parisi_1984}, reviewed and generalized recently in Ref.~\cite{Mazzola2017}. The auxiliary fields at the $n^{\rm th}$ Markov chain iteration are updated via the discretized Langevin equation:
\begin{equation}\label{eq:upFMD}
    \vec \sigma_{n+1} = \vec{\sigma}_{n} + \Delta_{\rm MD} S^{-1} \vec f_{n} + \sqrt{2\Delta_{\rm MD}} S^{-1/2} \vec z_{n}
\end{equation}
where $\vec{\sigma} = \{\bar{\sigma}_j^l \}$ are a shorthand notations for 
the fields represented as a $N \times T$ dimensional vector, $\Delta_{\rm MD}$ is the molecular-dynamics (MD) time step, $S_{l,m}^{i,j}$ is the acceleration matrix, that is chosen diagonal in the spatial indices  $S_{l,m}^{i,j} =P_{l,m} \delta_{i,j}$ and corresponding to  the harmonic classical part of the partition function,  $\vec z_n$ are normally distributed random vectors, and  $\vec f_n = \{ f_l^j\}_n$ are generalized forces with components: 
\begin{equation}
   \{ f_l^j \}_n = \partial_{\sigma_l^j} \ln(W(\{ \sigma_l^j \}_n)  
\end{equation}
It has been shown~\cite{assaad_md} that, within the complex auxiliary-field technique, the MD is free of ergodicity issues and we are able therefore to reproduce the results for the standard Hubbard model 
with $g=0$ (see Fig.~\ref{fig1}), which represents the most difficult case in our approach as the acceleration matrix $S$ turns out to be the trivial identity matrix. Better choices should be possible but this study is beyond the main purpose of this work.

\textit{Results: }
In order to access the information about the order parameters we examine the charge and spin correlations of the model for different values of Hubbard interaction and electron-phonon coupling on  $L\times L$ square lattices, at $\omega_0/t=1$. 

We adopt the recently proposed dynamic scaling~\cite{sandvik_scaling}.
We break the spin symmetry with 
the wavefunction $\Psi$, so that the antiferromagnetic order remains
in the $z$ direction for 
the chosen projection times $\beta t=L$. The thermodynamic limit $\beta t=L=\infty$ remains unbiased, whereas the finite-size results do not recover  
the singlet finite-$L$ ground state, reachable only 
for much larger projection times. This technique, 
has the considerable advantage to allow very stable simulations without the so called ``spikes'' (samples of correlation functions much far from their average values) implying infinite variance problems~\cite{zhang_variance}.
Within this set up $m_{\rm AF}$  can be computed as:
\begin{equation}
    m_{\rm AF} = \frac{1}{N} \sum_{i}{e^{i\textbf{Q} \cdot \textbf{r}_i}} \langle S_i \rangle \end{equation}
where $S_i=\frac{1}{2}(n_{i\uparrow}-n_{i\downarrow})$ is the value of the spin at site $i$ and the  
 charge structure factor is given by: 
\begin{equation}
    S_{\rm CDW}(\textbf{Q}) = \frac{1}{N} \sum_{i,j}{e^{i\textbf{Q}\cdot(\textbf{r}_i-\textbf{r}_j)}} \langle  n_i n_j \rangle 
\end{equation}
where $ \langle \cdots \rangle = { {\rm Tr}[ \exp( -\beta H/2) \cdots \exp( -\beta H/2)] \over \mathcal{Z} }$, and the pitching vector ${\bf Q}=(\pi,\pi)$. 
We consider the evolution of these quantities as a function of the  coupling $\lambda= {g^2 \over \omega_0}$. For $U \gg \lambda$ we have a Mott insulator with a finite antiferromagnetic (AF) order parameter $m_{\rm AF} >0$. 
As it is shown in Fig.~\ref{fig1}(b)
the dependence of this quantity on the 
MD time step $\Delta_{\rm MD}$ is rather smooth and can be safely extrapolated to the unbiased $\Delta_{\rm MD} \to 0$ limit.
Similar behavior is obtained for all the other quantities considered in this work.
The MD is particularly efficient just in the interesting region 
$\lambda \simeq U $ where phase transitions or at least 
competitions between antiferromagnetic, charge density wave or metallic and 
superconducting phases are expected \cite{Bauer2010}. In this case, 
the chosen acceleration matrix is particularly efficient because it  allows short correlation times $\simeq 1$ [see Fig.\ref{fig1}(b) and inset] and very weak time step dependency, allowing large-scale simulations in this region.

As far as the systematic error implied by a finite Trotter time $\Delta \tau$, this becomes negligible 
provided measurements are evaluated at the middle of the kinetic-energy propagator $\exp( -\Delta \tau K)$~\cite{note_Trotter},
 because in this way the error turns out quadratic in $\Delta \tau$. 
We have adopted $\Delta \tau t=0.1$ 
in all forthcoming calculations with an estimated error of less than $1 \%$ 
in all quantities studied.


\begin{figure}
  \begin{center}
    \includegraphics[width=.9\columnwidth]{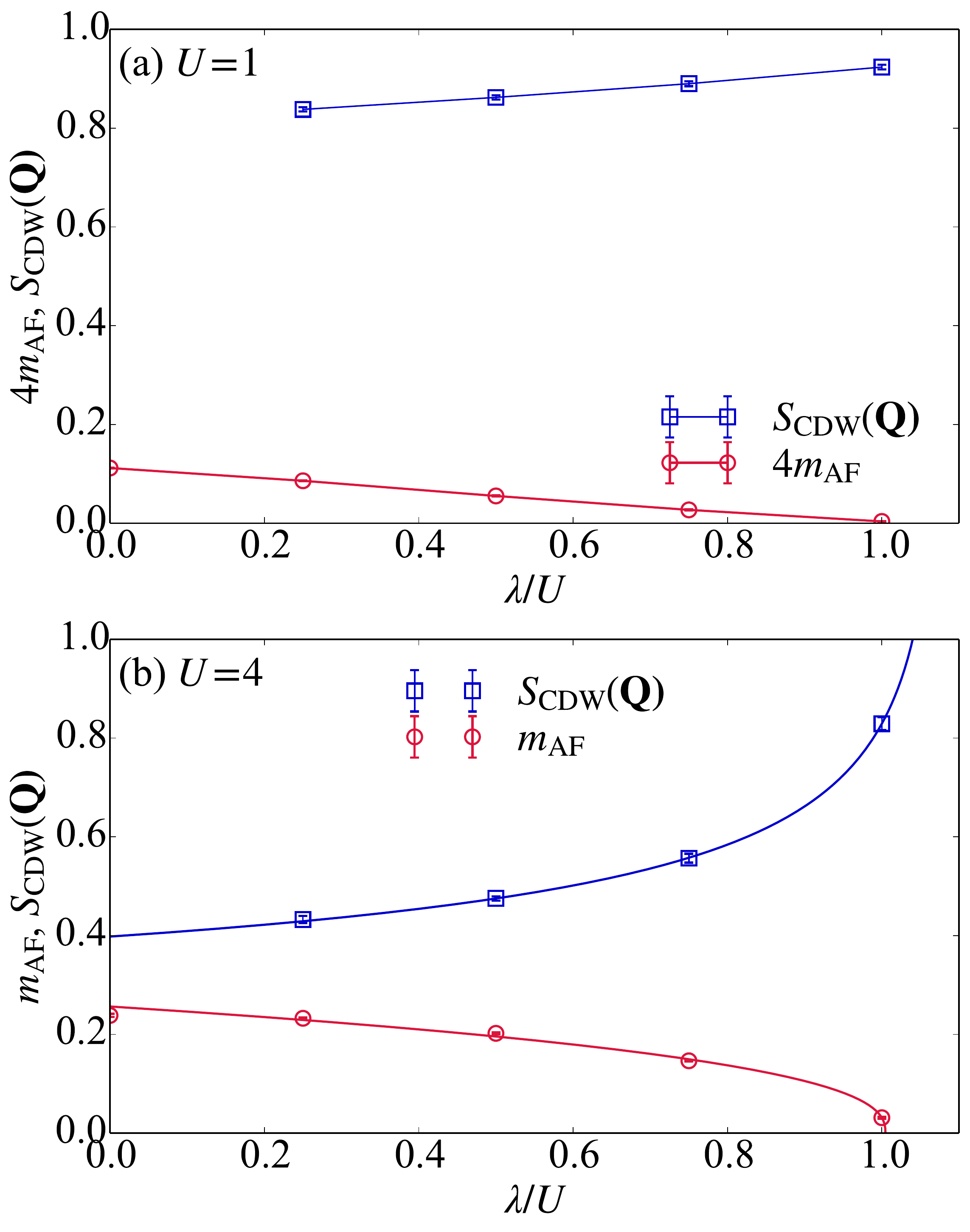}
    \caption{
      \label{fig4}
      Ground state properties of the Hubbard-Holstein model obtained by extrapolating the antiferromagnetic order parameters to the thermodynamic limit
      for (a) $U$=1 and (b) $U$=4. The solid lines in (b) are the fit to $m_{\rm AF}$ and 
      $S_{\rm CDW}(\bf{Q})$ of the form $\gamma\left(\frac{\lambda_{\rm c}-\lambda}{U}\right)^\theta$ with
      $\gamma$, $\theta$, and $\lambda_{\rm c}$ being the fitting parameters
      determined by the least-squares method. 
      The parameters are found to be 
      $\gamma=0.256(8)$, $\theta=0.39(5)$, and $\lambda_{\rm c}=\lambda_{\rm c}^{\rm AF}=4.02(2)$
      for $m_{\rm AF}$ and 
      $\gamma=0.407(2)$, $\theta=-0.286(12)$, and $\lambda_{\rm c}=\lambda_{\rm c}^{\rm CDW} = 4.33(4)$
      for $S_{\rm CDW}(\bf{Q})$. 
    }
  \end{center}
\end{figure}

We have performed a finite-size scaling of $m_{\rm AF}$ and $S_{\rm CDW}({\bf Q})$ 
for $U/t=4$ and $U/t=1$ using  clusters of size ranging from $6\times6$ to $18\times 18$ 
for several couplings $\lambda \le U$ 
and obtained 
the phase diagram reported in Fig.~\ref{fig4}.
As it is seen, the antiferromagnetic order drops continuously to much smaller values when we increase $\lambda$ and suggests a continuous transition to a non magnetic phase at $\lambda = \lambda_c^{\rm AF} \simeq U$. 
 Within this assumption, 
and considering that the pure Holstein-model for $U=0$ (i.e. $\lambda \gg U$) 
displays charge-density-wave (CDW) order,  $S_{\rm CDW}({\bf Q})$ should diverge for $ \lambda \to \lambda_c^{\rm CDW}$
from below. 
Quite interestingly, the results reported  in Fig.~\ref{fig4}, suggest  that $\lambda_c^{\rm CDW}$ 
is significantly larger than $\lambda_c^{\rm AF}$, because at small $U/t$ there is no evidence of the $S_{\rm CDW}({\bf Q})$ divergence,
whereas for $U/t=4$, despite the fit of the data are consistent with a very small critical exponent $\theta$,  $\lambda_c^{\rm CDW}$  
is about 8\% larger than $\lambda_c^{\rm AF}$. Indeed  if we fix $\theta$ to a larger value in the fit, we obtain an even larger value of $\lambda_c^{\rm CDW}$.


\textit{Conclusions: }
In this work we have presented an original way to get rid of the sign problem 
in the Hubbard-Holstein model by using an appropriate auxiliary-field transformation combined with an exact integration of the phonon degrees of freedom. The Hubbard-Holstein  model has been considered so far with algorithms  affected by the sign problem, because, though at half filling, does not satisfy the particle-hole transformation in the electronic degrees of freedom:
\begin{equation}
c^\dag_{j,\sigma} = (-1)^j c_{j,-\sigma}
\end{equation}
where $(-1)^j=1$ ($-1$) if $j$ belongs to the A (B) sublattice.
This transformation leaves unchanged the model without electron-phonon coupling but changes its sign when present. The key idea of this work is to employ an exact  integration of the phonon degrees of freedom, that  allow to recover this property and get rid of the sign problem, at least in a relevant parameter region $U> \lambda$. As we have shown the region $\lambda \simeq U$  is important 
because it is close to the phase transition of the model, and was previously 
inaccessible by numerical methods 
due to very severe sign problems~\cite{Johnston2013}.
On the other hand  in realistic materials the Coulomb energy is much larger than the electron-phonon coupling, as well as the phonon frequency $\omega_0$ 
and therefore the region $\lambda < U$, that can be studied 
with the present technique, is certainly the most important region for modelling realistic materials with the Hubbard-Holstein Hamiltonian.

Though the Hubbard-Holstein model is highly idealized, it is interesting to establish some benchmark results for the magnetic order parameter and the density structure factor (see Table~\ref{Table}). 
We see that our estimated $m_{\rm AF}$  compares well with the established benchmarks~\cite{SZhang2016} for $\lambda=0$, and  remains  approximately the same 
for $\lambda \ll U$, but with no evidence of CDW order, because $S_{\rm CDW}(\bf{Q})$ is clearly finite. Thus, as soon as 
$\lambda>0$, the electron-phonon coupling breaks the pseudo $SU(2)$ symmetry of the pure Hubbard model,
and kills the CDW, leaving  the AF order alone, in agreement with a rigorous theorem, recently proved \cite{Miyao2016}.  This feature reminds 
the phase diagram of the  negative $U$ Hubbard model where the CDW order disappears immediately  by a tiny amount of doping \cite{Scalettar1989}.

\begin{table}
  \caption{Values of $m_{\rm AF}$ and $S_{\rm CDW}(\textbf{Q})$ in the thermodynamic limits,
    for different values of $\lambda$, $U/t$ at $\omega_0=t$.\label{Table}}
  \begin{tabular}{lccccc}
    \hline
    \hline
    & \multicolumn{2}{c}{$U/t=1$}
    & &\multicolumn{2}{c}{$U/t=4$}
    \\
    \cline{2-3}
    \cline{5-6}
    $\lambda/U$ & $m_{\rm AF}$ & $S_{\rm CDW}(\bf{Q})$ & & $m_{\rm AF}$  & $S_{\rm CDW}(\bf{Q})$  \\
    \hline
    0  &0.0280(2) & --      & & 0.238(3) & --  \\
    0.25  &0.0215(3) &0.838(4) & & 0.232(2) & 0.433(7)  \\
    0.50  &0.0138(3) &0.862(4) & & 0.202(4) & 0.475(4)  \\
    0.75  &0.0068(4) &0.890(5) & & 0.146(2) & 0.557(9)  \\
    1  &0.0009(1) &0.924(5) & & 0.031(2) & 0.83(1)  \\
    \hline
    \hline
  \end{tabular}
\end{table}

Finally, since the transition to a non magnetic phase is very close to 
$\lambda \simeq U$, we have been able to determine some aspects of its phase diagram, namely that the transition is most likely continuous, at least up to $U/t=4$, 
as no evidence of a first order transition has been found for the $U$ values so far studied.
Also, rather unexpectedly, an intermediate phase $ \lambda_c^{\rm AF} \le \lambda \le \lambda_c^{\rm CDW}$ with no AF and CDW orders, appears rather robust and 
wide, in contrast with previous DMFT and VMC results.

This technique can be possibly extended to many other models, so far affected by the sign problem, and may open the way to tackle other important 
models where the particle-hole symmetry is not satisfied,  first among all the 
Hubbard model at finite doping. Though we do not expect that the sign problem 
in this model can be definitively removed, this work certainly suggests that 
the sign of the weight $W(\{\sigma_l^j\})$ can be very likely improved, being a property of the appropriate auxiliary field chosen, and the 
degrees of freedom selected in the path integral, where enormous freedom has not been so far explored.

\acknowledgments
S. Karakuzu acknowledges Prof. Richard Scalettar and Dr. Natanael C. Costa for useful discussions and providing data for comparison during the early stages of this work. We acknowledge useful discussions with Federico Becca. Computational resources were provided by CINECA.
\bibliography{sample}

\begin{thebibliography}{41}
\bibitem [{not({\natexlab{a}})}]{note_matA}%
  \BibitemOpen
  \href@noop {} {} ({\natexlab{a}}),\ \bibinfo {note} {strictly speaking in
  this case the diagonal elements $A_{l,l}$ for $l=1,T$ have to be divided by
  two, but this change does not allow a simple diagonalization of the matrix
  and on the other hand corresponds only to an irrelevant change of
  $|\Psi\rangle.$}\BibitemShut {NoStop}%
\bibitem [{not({\natexlab{b}})}]{note_Trotter}%
  \BibitemOpen
  \href@noop {} {} ({\natexlab{b}}),\ \bibinfo {note} {i.e. measurements are
  done by inserting the operator $O$ corresponding to a given correlation
  function, in the following way: $\cdots \exp{( -{\Delta \tau \over 2} K)} O
  \exp{( -{\Delta \tau \over 2} K )}\cdots$, where the dots indicate the
  remaining one body propagators in $W(\{ \sigma_l^j \})$}\BibitemShut
  {NoStop}%
\end{thebibliography}

\end{document}